# Extreme Value Theory for Time Series using Peak-Over-Threshold method

Gianluca Rosso

*GradStat – RSS, the Royal Statistical Society, London*
*Correspondent Researcher – SIS, Società Italiana di Statistica, Roma*
gianluca.rosso@sis-statistica.org

ABSTRACT: this brief paper summarize the chances offered by the Peak-Over-Threshold method, related with analysis of extremes. Identification of appropriate Value at Risk can be solved by fitting data with a Generalized Pareto Distribution. Also an estimation of value for the Expected Shortfall can be useful, and the application of these few concepts are valid for the most wide range of risk analysis, from the financial application to the operational risk assessment, through the analysis for climate time series; resolving the problem of borderline data.

In Extreme Value Theory (EVT) the method defined as Peak-Over-Threshold (POT) is common used. It is also commonplace to calculate its Value At Risk (VaR) and Expected Shortfall (ES) fitting data with a Generalized Pareto Distribution (GPD). This method is well known in financial risk analysis, but it is in general applicable to all kind of risk analysis (for example in climate time series; see Rick Katz, Extreme Value Analysis for Climate Time Series, Institute for Mathematics Applied to Geosciences National Center for Atmospheric Research Boulder, CO USA).

We know that the Normal Distribution is assumed as a standard for the most of analysis. The Normal Distribution has two parameters: mean and standard deviation.

$$N = (\mu, \sigma^2) \tag{1}$$

where

$$\mu = \frac{1}{n}\sum_{i=1}^{n} x_i \tag{2}$$

$$\sigma^2 = \frac{1}{n}\sum_{i=1}^{n}(x_i - \mu)^2 \tag{3}$$

Is assumed that the distribution in symmetrical, and that the mean divide the distribution into two symmetrical parts. The top of the distribution (corresponding with the mean) is defined by

$$\text{Inflection point} = \frac{1}{\sigma\sqrt{2\pi}} \tag{4}$$

The height of the density at any value x is given by

$$f(x) = \frac{1}{\sigma\sqrt{2\pi}}\, e^{-\frac{1}{2}\left(\frac{x-\mu}{\sigma}\right)^2} \tag{5}$$

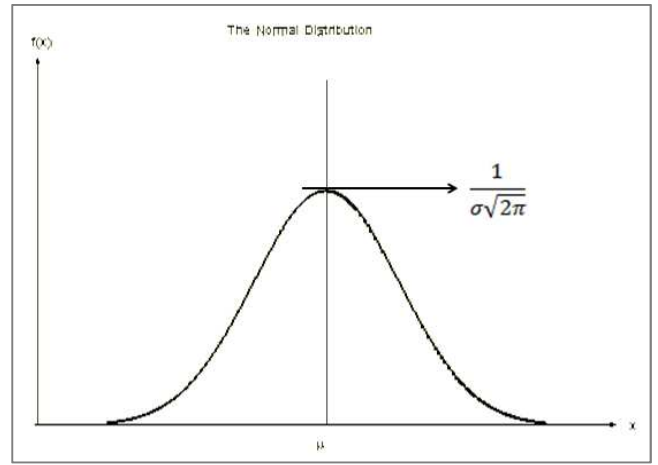

*Fig. 1*

When it is not easy to assume that the owned data could fitting a Normal Distribution, the way is to use a LogNormal Distribution. This function transform not-normal distribution into normal. Its shape is not symmetrical, therefore its inflection point and mean are not corresponding:

$$\text{Inflection point} = e^{\mu - \sigma^2} \tag{6}$$

$$\mu = \frac{1}{n}\sum_{i=1}^{n} \ln x_i \tag{7}$$

$$\sigma^2 = \frac{1}{n}\sum_{i=1}^{n}(\ln x_i - \mu)^2 \tag{8}$$

The height of the density at any value x is given by

$$f(x) = \frac{1}{x\sigma\sqrt{2\pi}}\, e^{-\frac{1}{2}\left(\frac{\ln x - \mu}{\sigma}\right)^2} \qquad x > 0 \tag{9}$$

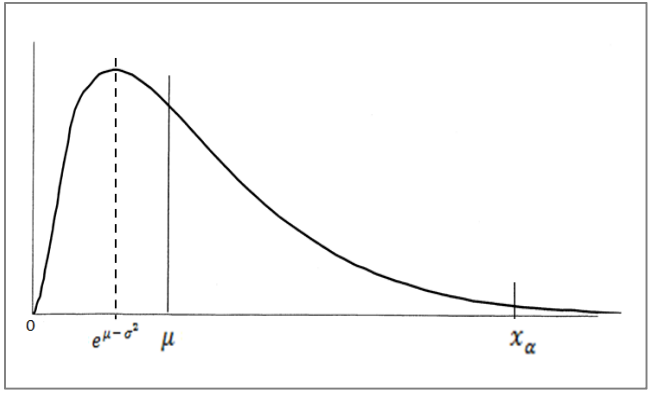

*Fig. 2*

Looking at the figure above, we can define

$$0 - \mu : \quad \int_0^{\mu} f(x)dx \quad \text{Expected Loss (EL)}$$

$$\mu - x_\alpha : \quad \int_\mu^{x_\alpha} f(x)dx \quad \text{Unexpected Loss (UL)}$$

$$x_\alpha - \infty : \quad \int_{x_\alpha}^{\infty} f(x)dx \quad \text{Worst Case (WC)}$$

(10, 11, 12)

The value of $x_\alpha$ is decided in according with the attended results of the analysis, or in according with rules defined by regulators. For example a very precuationally value for $\alpha$ could be at 99.5%.
Once decided the value for $\alpha$, also the value for $VaR_\alpha$ will be strictly linked to. The density function between 0 and $x_\alpha$ represents the VaR

$$VaR = EL + UL \tag{13}$$

$$VaR_\alpha = x_\alpha \tag{14}$$

The Expected Shortfall (ES) is

$$ES_\alpha = \frac{1}{\alpha} \int_{x_\alpha}^{\infty} f(x)dx \tag{15}$$

and therefore is

$$WC = \alpha\, ES_\alpha \tag{16}$$

So, as $VaR_\alpha = x_\alpha$ it is also true that

$$VaR_\alpha(X) = X_{(n\alpha)} \tag{17}$$

$$VaR_\alpha(X) = \inf\{x \in \mathbb{R} : P(X > x) \leq 1-\alpha\}$$
$$= \inf\{x \in \mathbb{R} : F_X(x) \geq \alpha\} \quad \alpha \in (0,1) \tag{18}$$

This means that the $VaR_\alpha$ is given by the smallest value $x$ such that the actual loss $X$ exceeds $x$ with the probability $1-\alpha$, at most.
Note that in this case we have the Worst Case that lies not more between the given $\alpha$ and $\infty$, but (probabilistically speaking) between $\alpha$ and 1.

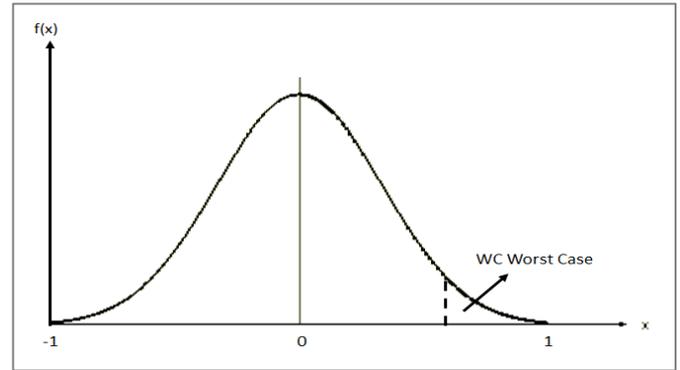

*Fig. 3*

So, our Worst Case will be

$$WC = 100\,(1-\alpha)\% \tag{19}$$

and $ES_\alpha$ will be

$$ES_\alpha(X) = \frac{1}{1-\alpha} \int_\alpha^1 VaR_\gamma(X)\, d\gamma \tag{20}$$

The model will be defined by the function

$$f(x) = f(x_1, x_2, \ldots, x_n | \theta)$$
$$= f(x_1|\theta) \cdot f(x_2|\theta) \cdot \ldots \cdot f(x_n|\theta) \tag{21}$$

where $\theta$ are the parameters of the model.
As said above

$$f(x) = \prod_{i=1}^{n} f(x_i | \theta) \tag{22}$$

in log-normality is

$$f(x) = \sum_{i=1}^{n} ln f(x_i \mid \theta)$$
(23)

Now, we can consider that one of the best way to analyze the peak of our time series is the POT method. About the method we suggest to refer to the very large literature written during last years. We focus now to the analysis via GPD and the possible way to estimate VaR and ES.
Let $X_i$, $i = 1, \ldots, N$ our data set (or sample), and let $u$ the value of the threshold, so that

$$y = x - u$$
(24)

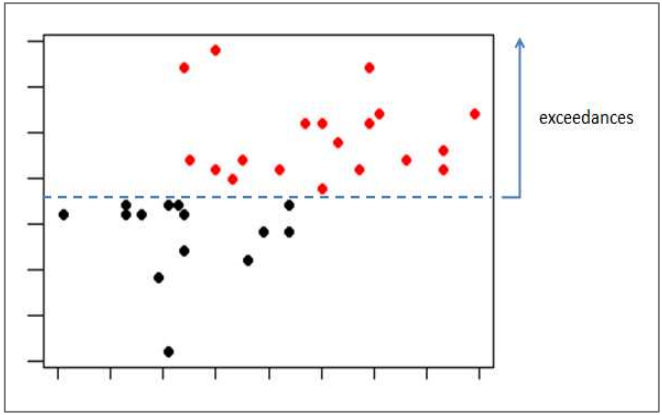

*Fig. 4*

The GPD is defined by

$$G_{\xi,\beta}(x) = \begin{cases} 1 - \left(1 + \frac{\xi x}{\beta}\right)^{-\frac{1}{\xi}} & if \ \xi \neq 0 \\ 1 - exp\left(-\frac{x}{\beta}\right) & if \ \xi = 0 \end{cases}$$
(25)

and are valid

$$\beta > 0 \ and \ x \geq 0 \ \rightarrow \ \xi \geq 0$$

$$0 \leq x \leq -\frac{\beta}{\xi} \quad \rightarrow \ \xi < 0$$
(26)

where $\xi$ is the shape parameter, and $\beta$ is the scale parameter. So, we need to put our exceedances in relation with the GPD

$$F_u(x) = P(X - u \leq x \mid X > u) = \frac{F(x + u) - F(u)}{1 - F(u)}$$
(27)

since

$$F_u(y) \approx G_{\xi,\beta}(y)$$
(28)

it is that

$$F(x) = \big(1 - F(u)\big) \cdot G_{\xi,\beta}(x - u) + F(u) \quad for \ x > u$$
(29)

where

$$F(u) = \frac{N - m}{N}$$
(30)

given $N$ the number of observations and $m$ the number of exceedances.
So the estimation is

$$\widehat{F(x)} = 1 - \frac{m}{N}\left(1 + \frac{\hat{\xi}(x - u)}{\hat{\beta}}\right)^{-\frac{1}{\hat{\xi}}}$$
(31)

We know that observations are denoted by $x$, so for $x$ and $VaR_\alpha$ is

$$x = VaR_\alpha$$
(32)

$$\alpha = 1 - \frac{m}{N}\left(1 + \frac{\hat{\xi}(\widehat{VaR_\alpha} - u)}{\hat{\beta}}\right)^{-\frac{1}{\hat{\beta}}}$$
(33)

$$\widehat{VaR_\alpha} = u + \frac{\hat{\beta}}{\hat{\xi}}\left\{\left[\frac{N}{m}(1 - \alpha)\right]^{-\hat{\xi}} - 1\right\}$$
(34)

About the Worst Case, and reguarding what said above, it is possibile to define the estimation for ES

$$WC = \propto ES_\propto$$
(35)

$$\widehat{ES_\propto} = \frac{1}{1 - \propto} \int_\propto^1 \widehat{VaR_y} \ dy = \frac{\widehat{VaR_\propto}}{1 - \hat{\xi}} + \frac{\hat{\beta} - \hat{\xi}}{1 - \hat{\xi}}$$
(36)

where

| | |
|---|---|
| $\hat{\xi}, \hat{\beta}$ | = estimated GPD parameters |
| N | = number of observations |
| m | = number of exceedances |
| u | = value of the threshold |

Brief conclusions.

The POT method assumes that once we have placed a threshold, all the items over this level are peaks. But as I wrote in a paper about the clustering (Rosso, 2014, Outliers Emphasis on Cluster Analysis - The use of squared Euclidean distance and fuzzy clustering to detect outliers in a dataset, arXiv:1403.5417), why items very close between them must be attributed to a cluster instead to another one? The conclusion of that paper was that we must have a method that runs more close with the fuzzy thinking of our brain. The solution was: fuzzy clustering, in first instance. After that, possibilistic clustering could be a substantial help to manage data with a more flexible vision.

Take a look at Figure 4, again, here down as Figure 5. We should consider the item number 1 a peak, and the item number 2 should be considered as a standard data. This is although they are so close in terms of value of $x$.

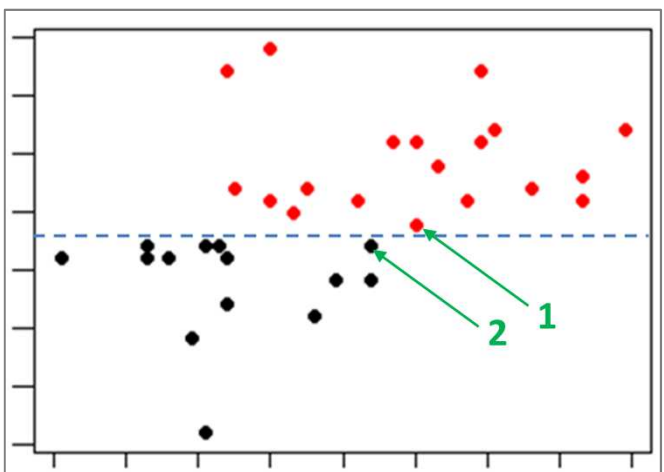

*Fig. 5*

It will be absolutely more easy to consider a plot as in Figure 6.

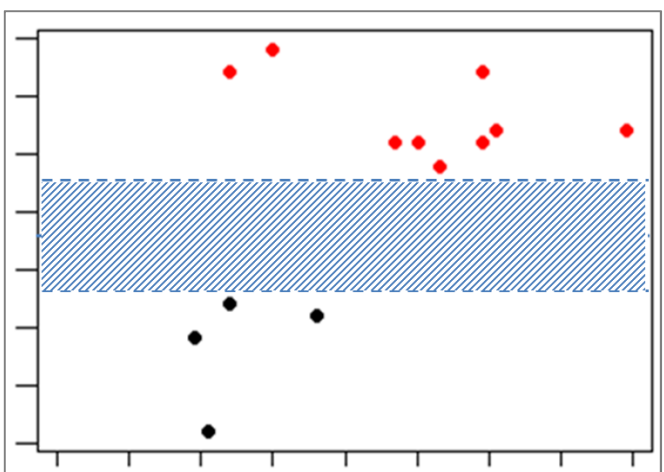

*Fig. 6*

Surely in this situation we have no dubt about peaks. We have a large blue zone without data, virtually an extremely wide threshold. But in plot in Figure 5 we must mitigate the euclidean distance between the threshold line and each single item, or event. We need an average value, but thought in term of risk because we must remember that our analysis is for the value at risk. So, the applications of the rules described above can conduct to an assessment numerically substainable, and surely consistent in its reliability, despite borderline cases.

———————————————